\documentclass[aps,pre,preprint,showpacs]{revtex4}
\usepackage{graphicx}

\begin{document}

\title{The effect of the range of interaction on the phase diagram of a
globular protein}
\author{James F. Lutsko and Gr\'{e}goire Nicolis}
\affiliation{Center for Nonlinear Phenomena and Complex Systems, Universit\'{e} Libre de
Bruxelles, C.P. 231, Blvd. du Triomphe, 1050 Brussels, Belgium}
\pacs{87.15.Aa,81.30.Dz,05.20.Jj}

\begin{abstract}
Thermodynamic perturbation theory is applied to the model of globular proteins
studied by ten Wolde and Frenkel (Science 277, pg. 1976) using computer
simulation. It is found that the reported phase diagrams are accurately
reproduced. The calculations show how the phase diagram can be tuned as a
function of the lengthscale of the potential.
\end{abstract}

\maketitle

\section{Introduction}

One of the most important problems in biophysics is the characterization of
the structure of proteins. The experimental determination of protein
structure by means of x-ray diffraction requires that the proteins be
prepared as good quality crystals which turns out to be difficult to
achieve. Given the fact that recent years have seen an explosion in the
number of proteins which have been isolated, the need is therefore greater
than ever for efficient methods to produce such crystals. Without finely tuned experimental
conditions, often discovered through laborious trial and error,
crystallization may not occur on laboratory time-scales or amorphous, rather
than crystalline, structures may form. The recent observation by ten Wolde
and Frenkel\cite{tWF} of enhanced nucleation of a model protein in the
vicinity of a metastable critical point is thus of great interest and could
lead to more efficient means of crystallization if such conditions can be
easily identified for a given protein.

Wilson noted that favorable conditions for crystallization are correlated
with the behavior of the osmotic second virial coefficient\cite{GeorgeWilson}
and, hence, depend sensitively on temperature. If the second virial
coefficient is too large, crystallization occurs slowly and if it is too
small, amorphous solids form. By comparing the experimentally determined
precipitation boundaries for several different globular proteins as a
function of interaction range, controlled by means of the background ionic
strength, Rosenbaum et al have shown that the phase diagrams of a large
class of globular proteins can be mapped onto those of simple fluids
interacting via central force potentials consisting of hard cores and
short-ranged attractive tails\cite{Rosenbaum1},\cite{Rosenbaum2}. They also
discuss the important fact that the range of interaction can be tuned by
varying the composition of the solvents used. The attraction must, in
general, be very short ranged if this model is to apply since a
fluid-fluid phase transition is not typically observe experimentally\cite%
{Rosenbaum2} and it is known that this transition is only suppressed in
simple fluids when the attractions are very short ranged\cite{FrenkelYukawa}%
. These studies therefore support the conclusion that the study of simple
fluids interacting via potentials with short-ranged attractive tails can
give insight into nucleation of the crystalline phase of a large class of
globular proteins. ten Wolde and Frenkel have studied nucleation
of a particular model globular protein consisting of a hard-core and a
modified Lennard-Jones tail by direct free energy measurements obtained from
computer simulations\cite{tWF}. They found that the nucleation rate of a
stable FCC solid phase could be significantly enhanced in the vicinity of a
metastable critical point. The enhancement is due to the possibility that a
density fluctuation in the vapor phase is able to first nucleate a
metastable droplet of denser fluid which, in turn, forms a crystal nucleus.
The fact that intermediate metastable states can accelerate barrier crossing
has been confirmed using kinetic models\cite{Nicolis1} and the physics of
the proposed non-classical nucleation model has also been confirmed by
theoretical studies based on density functional models\cite{OxtobyProtein},%
\cite{GuntonProtein}. This observation opens up the possibility of efficiently producing
good quality protein crystals provided that it is understood how to tune the
interactions governing a given protein so that its phase diagram possesses
such a metastable state under experimentally accessible conditions. A
prerequisite for achieving this is to go beyond the heavily parameterized
studies conducted so far and to be able to accurately predict phase diagrams
given knowledge of the range of the protein interactions. In this paper, we
describe the application of thermodynamic perturbation theory to calculate
the phase diagram based solely on the interaction model. In so far as the
range of interaction is important, and not the detailed functional forms,
this approach, if successful, gives a direct connection between the phase
diagram and the range of interaction without the need for further,
phenomenological parameterizations. We show that the theory can be used to
successfully reproduce the phase diagrams of ten\ Wolde and Frenkel based
only on the interaction potential and assess the effect of the range of the
interatomic potential on the structure of the phase diagram. 

In the next
Section, the formalism used in our calculations is outlined. This involves
the standard Weeks-Chandler-Andersen perturbation theory with modifications
to improve its accuracy at high densities. The third Section discusses the
application of the perturbation theory to the ten Wolde-Frenkel interaction
model. Whether or not perturbation theory is applicable to this type of
system is not immediately evident: the hard-core square well potential has
long served as a test case for developments in perturbation theory\cite%
{BarkerHend_WhatIsLiquid}. So we show how the size of the various
contributions to the total free energy varies with temperature and that
second order contributions to the free energy are of negligible importance.
In the fourth Section, the calculated phase diagram for the  hard core
plus modified Lennard-Jones tail is shown to be in good agreement with the
reported Monte Carlo (MC)\ results. Since the perturbation theory is also
well known\cite{Ree2} to give a good description of long-ranged potentials
such as the standard Lennard-Jones, we expect that it can be used with some
confidence to explore the effect of the length scale of the potential on the
phase behavior of the systems. To illustrate, we present the phase diagram
as a function of the range of the modified Lennard-Jones tail and show that
the appearance of the metastable state requires only a minor modification of
the range of the potential. The final Section contains our conclusions where
we discuss the prospect for using the perturbation theory free energy
function as the basis for density functional studies of the nucleation
process and for studies of the effect of fluctuations on the nucleation rate.

\section{Thermodynamic perturbation theory}

Thermodynamic perturbation theory allows one to express the Helmholtz free
energy $F$ of a system in terms of a perturbative expansion about some
reference system. There are a number of different approaches to constructing
the perturbative expansions such as the well known Weeks-Chandler-Andersen
(WCA)\cite{WCA1},\cite{WCA2},\cite{WCA3},\cite{HansenMcdonald} theory and
the more recent Mansoori-Canfield/Rasaiah-Stell theory\cite{Stell2004}. The
latter appears to be more accurate for systems with soft repulsions at small
separations while the former works better for systems with stronger
repulsions. Here, we will be interested in a hard-core potential with a
modified Lennard-Jones tail, so we use the WCA theory as modified by Ree and
coworkers\cite{Ree1},\cite{Ree2},\cite{ReeSol} as discussed below. The first
step is to divide the potential into a (mostly) repulsive short-ranged part
and a longer ranged (mostly) attractive tail according to the prescription 
\begin{eqnarray}
v(r) &=&v_{0}\left( r\right) +w\left( r\right)  \label{break} \\
v_{0}\left( r\right) &=&\left\{ 
\begin{array}{c}
v\left( r\right) -v\left( r_{0}\right) -v^{\prime }\left( r_{0}\right)
\left( r-r_{0}\right) ,\;r<r_{0} \\ 
0,\;r>r_{0}%
\end{array}%
\right.  \nonumber \\
w\left( r\right) &=&\left\{ 
\begin{array}{c}
v\left( r_{0}\right) +v^{\prime }\left( r_{0}\right) \left( r-r_{0}\right)
,\;r<r_{0} \\ 
v\left( r\right) ,\;r>r_{0}%
\end{array}%
\right. .  \nonumber
\end{eqnarray}%
The short ranged part is generally repulsive and can therefore be well
approximated by a hard-sphere reference system. The long-ranged tail
describes the attractive forces and must also be accounted for so that
distinct liquid and gas phases exist (i.e. so that the phase diagram
exhibits a Van der Waals loop). There are a number of versions of the
WCA-type perturbation theory depending on the choice of the separation point 
$r_{0}$. Barker and Henderson\cite{BarkerHend} chose the separation point $%
r_{0}$ to be the point at which the potential goes to zero, $v\left(
r_{0}\right) =0$, (they also did not include the linear term in the
expressions above). Subsequently, WCA achieved a better description of the
Lennard-Jones phase diagram by taking the separation point to be at the
minimum of the potential, $v^{\prime }\left( r_{0}\right) =0$ . Ree\cite%
{Ree1} first suggested that the free energy be minimized with respect to $%
r_{0},$ and introduced the linear terms in eq.(\ref{break}), in which case
the first-order perturbation theory is equivalent to a variational theory
based on the Gibbs-Bugolyubov inequalities\cite{HansenMcdonald}. Later, Ree
and coworkers showed that essentially the same results could be achieved
with the prescription%
\begin{equation}
r_{0}=\min \left( r_{\min },r_{nn}\right)  \label{break1}
\end{equation}%
where $r_{\min }$ is the minimum of the potential, $v^{\prime }\left(
r_{\min }\right) =0$ and  $r_{nn}=2^{\frac{1}{6}}\rho ^{-1/3}$, where $\rho $ is
the density, is the FCC nearest-neighbor distance\cite{Ree2}. For low
densities, this amounts to the original WCA prescription whereas for higher
densities, the separation point decreases with increasing density. In this
case, the linear term in the definition of $v_{0}\left( r\right) $ ensures
the continuity of the first derivative of the potential. Calculations for
the Lennard-Jones potential, as well as inverse power potentials, show that
this modification of the original WCA theory gives improved results at high
density. Finally, eq.(\ref{break1}) was modified to switch smoothly from $%
r_{\min }$ to $r_{nn}$ as the density increases so as to avoid
discontinuities in the free energy as a function of density and thus 
singularities in the pressure\cite{ReeSol}. Below, we will refer to this
final form of the Weeks-Chandler-Andersen-Ree theory as the WCAR theory.

\subsection{Contribution of the long-ranged part or the potential}

The contribution of the long-ranged part of the potential to the free energy
is handled perturbatively in the so-called high-temperature expansion\cite%
{HansenMcdonald}\ 
\begin{equation}
\frac{1}{N}\beta F - \frac{1}{N}\beta F_{0} = \frac{1}{N}\beta \left\langle
W\right\rangle _{0}+\frac{1}{2N}\beta ^{2}\left[ \left\langle
W^{2}\right\rangle _{0}-\left\langle W\right\rangle _{0}^{2}\right] +...
\end{equation}%
where $F_{0}$ is the free energy of a system of $N$ particles subject only
to the short-ranged potential $v_{0}\left( r\right) $ at inverse temperature 
$\beta =1/k_{B}T$ and where the total attractive energy is 
\begin{equation}
W=\sum_{1\leq i<j\leq N}w\left( r_{ij}\right) .
\end{equation}%
The brackets $\left\langle ...\right\rangle _{0}$ indicate an equilibrium
average over a system interacting with the potential $v_{0}$. The first term
on the right is easily calculated since it only involves the pair
distribution function of the reference system%
\begin{equation}
\frac{1}{N}\beta \left\langle W\right\rangle _{0}=\frac{1}{2}\beta \rho \int
d\overrightarrow{r}\;g_{0}\left( r\right) w\left( r\right)
\end{equation}%
where $g_{0}\left( r\right) $ is the pair distribution function of the
reference system. The second term requires knowledge of three- and four-body
correlations for which good approximations are not available. Its value is
typically estimated using Barker and Henderson's ''macroscopic
compressibility'' approximation\cite{BHMacroscopicCompressibility},\cite{BarkerHend_WhatIsLiquid}%
\begin{equation}
\frac{1}{2N}\beta ^{2}\left[ \left\langle W^{2}\right\rangle
_{0}-\left\langle W\right\rangle _{0}^{2}\right] \simeq -\frac{1}{4}\beta
^{2}\rho \left( \frac{\partial \rho }{\partial \beta P_{0}}\right) \int d%
\overrightarrow{r}\;g_{0}\left( r\right) w^{2}\left( r\right) \label{macrocompress} 
\end{equation}%
where $P_{0}$ is the pressure of the reference system at temperature $%
k_{B}T=1/\beta $ and density $\rho $.

\subsection{Contribution of the short-ranged part of the potential}

The description of the reference system is again accomplished by
perturbation theory. Since the potential $v_{0}\left( r\right) $ is not very
different from a hard core potential, this perturbation theory does not
involve the high temperature expansion but, rather, involves a functional
expansion in the quantity $\exp \left( -\beta v_{0}\left( r\right) \right)
-\exp \left( -\beta v_{hs}\left( r;d\right) \right) $ where $v_{hs}\left(
r;d\right) $ is the hard sphere potential for a hard-sphere diameter $d$.
The result is%
\begin{equation}
\frac{1}{N}\beta F_{0}-\frac{1}{N}\beta F_{hs}\left( \rho d^{3}\right) = \int
d\overrightarrow{r}\;y_{hs}\left( r\right) \left( \exp \left( -\beta
v_{0}\left( r\right) \right) -\exp \left( -\beta v_{hs}\left( r;d\right)
\right) \right) +...  \label{awc}
\end{equation}%
where $y_{hs}\left( r,\rho d^{3}\right) $ is the hard-sphere cavity function,
 related to the pair distribution function through $g_{hs}(r)=
\exp \left( -\beta v_{hs}\left( r;d\right) \right)y_{hs}(r)$. Several
methods of choosing the hard-sphere diameter of the reference system are
common. The WCA prescription is to force the first order term to vanish%
\begin{equation}
0=\int d\overrightarrow{r}\;y_{hs}\left( r\right) \left( \exp \left( -\beta
v_{0}\left( r\right) \right) -\exp \left( -\beta v_{hs}\left( r;d\right)
\right) \right) .  \label{WCAdiameter}
\end{equation}%
and a simple expansion about $r=d$ gives the cruder Barker and Henderson
approximation\cite{HansenMcdonald} which gives%
\begin{equation}
\int dr\left( \exp \left( -\beta v_{0}\left( r\right) \right)
-1\right) +\int dr\left( 1-\exp \left( -\beta v_{hs}\left( r;d\right)
\right) \right) \simeq 0 .
\end{equation}%
In either case, one can then consistently approximate the pair distribution
function of the reference state as either 
\begin{equation}
g_{0}\left( r\right) \simeq g_{hs}\left( r\right)  \label{g1}
\end{equation}%
or%
\begin{equation}
g_{0}\left( r\right) \simeq \exp \left( -\beta v_{0}\left( r\right) \right)
y_{hs}\left( r\right)  \label{g2}
\end{equation}%
where the difference between using one expression or the other is of the
same size as neglected terms in the perturbation theory. Here, we follow Ree
et al\cite{Ree2} in using the WCA prescription for the hard-sphere diameter,
eq.(\ref{WCAdiameter}) and the first approximation, eq.(\ref{g1}), for the
pair distribution function. Then, the complete expression for the free
energy becomes%
\begin{eqnarray}
\frac{1}{N}\beta F &=&\frac{1}{N}\beta F_{hs}\left( \rho d^{3}\right) 
+\int d\overrightarrow{r}\;y_{hs}\left( r\right) \left( \exp \left( -\beta
v_{0}\left( r\right) \right) -\exp \left( -\beta v_{hs}\left( r;d\right)
\right) \right)  \label{fe} \\
&&+\frac{1}{2}\beta \rho \int d\overrightarrow{r}\;g_{hs}\left( r\right)
w\left( r\right)  \nonumber \\
&&-\frac{1}{4}\beta \rho \left( \frac{\partial \rho }{\partial P_{0}}\right)
\int d\overrightarrow{r}\;g_{hs}\left( r\right) w^{2}\left( r\right) . 
\nonumber
\end{eqnarray}%
The pressure, $P$, and chemical potential $\mu $ are calculated from the
free energy using the standard thermodynamic relations%
\begin{eqnarray}
\frac{\beta P}{\rho } &=&\rho \frac{\partial }{\partial \rho }\frac{1}{N}\beta F
\label{Pmu} \\
\beta \mu &=&\frac{1}{N}\beta F+\frac{\beta P}{\rho }.  \nonumber
\end{eqnarray}

\subsection{Description of the reference liquid}

The calculation of liquid phase free energies require as input the
properties of the hard sphere liquid. These are known to a high degree of
accuracy and introduce no significant uncertainty, nor any new parameters,
into the calculations.

The properties of low density hard-sphere liquids are well described by the
Percus-Yevick (PY) approximation but this is not adequate for the dense
liquids to be considered here. So for the hard sphere cavity function, we
have used the model of Henderson and Grundke\cite{HendGrundke} which
modifies the PY description so as to more accurately describe dense liquids.
The corresponding pair distribution function is then that of Verlet and Weiss%
\cite{VWgr} and the equation of state, as obtained from it by both the
compressibility equation and the pressure equation, is the Carnahan-Starling
equation of state\cite{HansenMcdonald}. The free energy as a function of
density follows immediately and is given by 
\begin{equation}
\frac{1}{N}\beta F_{hs}\left( \rho d^{3}\right) =\ln \left( \rho \Lambda
^{3}\right) -1+\eta \frac{4-3\eta }{\left( 1-\eta \right) ^{2}}
\end{equation}%
where $\eta =\rho d^{3}$. The second term of eq.(\ref{fe}) is easily
evaluated numerically as its kernel is sharply peaked about $r=d$ . The most
troublesome part of the calculation is the evaluation of the contributions
of the long-ranged part of the potential, $w\left( r\right) $ . One method
is to divide the necessary integrals along the lines%
\begin{equation}
\int d\overrightarrow{r}\;g_{hs}\left( r\right) w\left( r\right) =\int d%
\overrightarrow{r}\;w\left( r\right) +\int d\overrightarrow{r}\;\left(
g_{hs}\left( r\right) -1\right) w\left( r\right)
\end{equation}%
where the first piece can be calculated analytically and the second involves
the structure function $g_{hs}\left( r\right) -1$ which is relatively short
ranged allowing a numerical evaluation. However, at high densities this can
still be difficult to evaluate as the hard-sphere structure extends for
considerable distances. In the Appendix, we discuss a more efficient method
of evaluation based on Laplace transform techniques. We have used both
methods and obtained consistent results:\ in general, the second is much
easier to implement and numerically more stable.

\subsection{Description of the reference solid}

To calculate the properties of the solid phase, the same expressions are
used except that the reference free energy is now that of the hard-sphere
solid and the pair distribution function is the spherical average of the
hard-sphere pair distribution function. Both of these quantities can be
obtained by means of classical density functional theory, but here we choose
the simpler, and older, approach which makes use of analytic fits to 
the results of computer simulations together with the known high-density
limit of the equation of state. This limits the present calculations to the
investigation of the FCC solid phase as this is the only one for which
extensive simulations have been performed. We stress that
these fits are very good and that they introduce no new parameters into the
calculations of the phase diagrams.

In the calculations presented below, we have used the equation of state
proposed by Hall\cite{HallHsSolidEOS} 
\begin{eqnarray}
\frac{\beta P}{\rho } &=&3\frac{\eta }{\eta _{c}-\eta }%
+2.557696+0.1253077b+0.1762393b^{2}-1.053308b^{3} \\
&&+2.818621b^{4}-2.921934b^{5}+1.118413b^{6}  \nonumber \\
b &=&4\left( 1-\frac{\eta }{\eta _{c}}\right)   \nonumber
\end{eqnarray}%
where $\eta _{c}=\frac{\pi }{6}\sqrt{2}$ is the value of the packing
fraction at close packing. Notice that the first term is the high density
limit of the Lennard-Jones-Devonshire\ cell theory which is expected to be
exact near close packing (see, e.g., the discussion in \cite{AlderHsSolidEOS}%
). The free energy is then calculated by integrating from the desired
density to the close-packing density giving 
\begin{equation}
\beta F=\beta F_{0}-\int_{\eta }^{\eta _{c}}\left[ \frac{\beta P}{\rho }%
-\left( \frac{\beta P}{\rho }\right) _{LJD}\right] \frac{d\eta }{\eta }.
\end{equation}

For the spherically-averaged pair distribution function for the FCC solid,
we use the analytic fits of Kincaid and Weiss\cite{WKSolidGr} 
\begin{equation}
g_{KW}\left( r\right) =\left( A/x\right) \exp \left[ -w_{1}^{2}\left(
x-x_{1}\right) -w_{2}^{4}\left( x-x_{1}\right) ^{4}\right] +\frac{w}{24\eta 
\sqrt{\pi }}\sum_{i=2}^{\infty }\frac{n_{i}}{x_{i}x}\exp \left( -w^{2}\left(
x-x_{1}\right) ^{2}\right).
\end{equation}%
Here $x=r/d$, the parameter $A$ is fixed by requiring that the pressure
equation reproduce the Hall equation of state%
\begin{equation}
\left( \frac{\beta P}{\rho }\right) _{Hall}=1+4\eta g_{KW}\left( 1\right) ,
\end{equation}%
and the parameters $w_{1},$ $w_{2}$ and $w$ are given as functions of density by
analytic fits to the MC data\cite{WKSolidGr}. No such fit is given for the
parameter $x_{1}$ so its value must be determined by interpolating from the
values extracted from the MC data as given in \cite{WKSolidGr}. The
quantities $n_{i}$ and $x_{i}$ are the number of neighbors and the position
of the $i$-th lattice shell respectively. Note that Ree et al suggest using
the earlier parameterization of Weis\cite{WeissSolidGr} at lower densities,
where it is slightly more accurate, and the Kincaid-Weis version at higher
densities. We have not done this because it leads to discontinuities in the
free energy as a function of density at the point the switch is made. Since
these are just empirical fits, we do not believe there is a significant loss
of accuracy.

\bigskip

\section{Application to a model protein intermolecular potential}

\subsection{The potential}

The only input needed for the
perturbative calculation outlined in Section II is the intermolecular potential: there are no
phenomenological parameters to specify. 
The goal of this work is to show how to construct a realistic free energy functional
with which to study nucleation of protein crystallization using the model
potential of ten Wolde and Frenkel\cite{tWF}. This interaction model
consists of a hard-sphere pair potential with an attractive tail 
\begin{equation}
v\left( r\right) =\left\{ 
\begin{array}{c}
\infty \;,\;r<\sigma \\ 
\frac{4\varepsilon }{\alpha ^{2}}\left[ \frac{1}{\left( \left( \frac{r}{%
\sigma }\right) ^{2}-1\right) ^{6}}-\alpha \frac{1}{\left( \left( \frac{r}{%
\sigma }\right) ^{2}-1\right) ^{3}}\right] \;,\;r>\sigma%
\end{array}%
\right.
\end{equation}%
The tail is actually a modified Lennard-Jones potential and
the two are related by%
\begin{equation}
\Theta \left( r-\sigma \right) v\left( r\right) =\Theta \left( r-\sigma
\right) v_{LJ}\left( \alpha ^{1/6}\sigma \sqrt{\left( \frac{r}{\sigma }%
\right) ^{2}-1}\right) .
\end{equation}%
As such, the potential decays as a power law and is not short-ranged in the
usual sense. Nevertheless, as $\alpha $ becomes larger, the range of the
potential decreases: for example, the minimum of the potential is 
\begin{eqnarray}
\frac{r_{\min }}{\sigma } &=&\sqrt{1+\left( \frac{2}{\alpha }\right) ^{1/3}}
\\
v\left( r_{\min }\right) &=&-\varepsilon .  \nonumber
\end{eqnarray}%
which approaches the hard core for large $\alpha $ . Furthermore, for a
fixed position $r>r_{\min }$, the value of the potential decreases with
increasing $\alpha $ relative to its minimum. For example,%
\begin{equation}
v\left( 2\sigma \right) /v\left( r_{\min }\right) =\frac{108\alpha -4}{%
729\alpha ^{2}}
\end{equation}%
so that as $\alpha $ increases, the interactions of particles separated by
much more than $r_{\min }$ contribute less and less to the total energy
compared to the contribution of particles that are close together. Figure 1
shows the evolution of the shape of the potential as $\alpha $ increases.
The range of the potential varies from about $2.5$ hard sphere diameters for 
$\alpha =1$ to less than $1.25$ diameters for $\alpha =50$. Also shown in
the figure is the separation of the potential into long- and short-ranged
pieces for the case $\alpha =50$ where it is clear that even for this very
short-ranged potential, the long-ranged function $W(r)$ varies relatively
slowly compared to the short-ranged repulsive potential $V_{0}\left(
r\right) $.

\begin{figure*}[tbp]
\includegraphics[angle=0,scale=0.4]{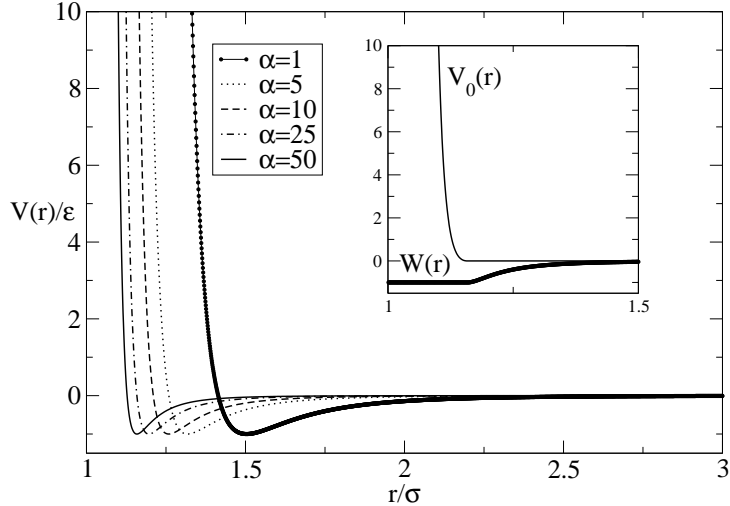}
\caption{The ten Wolde-Frekel potential as a function of $\protect\alpha$.
The inset shows the division of the potential into long-ranged and
short-ranged parts for $\protect\alpha=50$.}
\label{fig1}
\end{figure*}

\subsection{Comparison of various approximations}

Figure 2 shows the contributions of the various terms contributing to the
free energy at two temperatures. In both cases, the second order term is
seen to be negligible. This is because at low density, the free energy is
dominated by the ideal-gas contribution, all other contributions going to
zero with the density, whereas at moderate to high density, the
compressibility controlling the size of the contribution of the second order
term, see eq.(\ref{fe}), diminishes quickly from its zero density limit of $%
1.0$ to something on the order of $0.1$ at moderate densities and is of
order $0.01$ at high densities. We conclude that the second order
contributions, at least calculated within the macroscopic compressibility approximation,
 eq.(\ref{macrocompress}),
can be neglected.

\begin{figure*}[tbp]
\includegraphics[angle=0,scale=0.4]{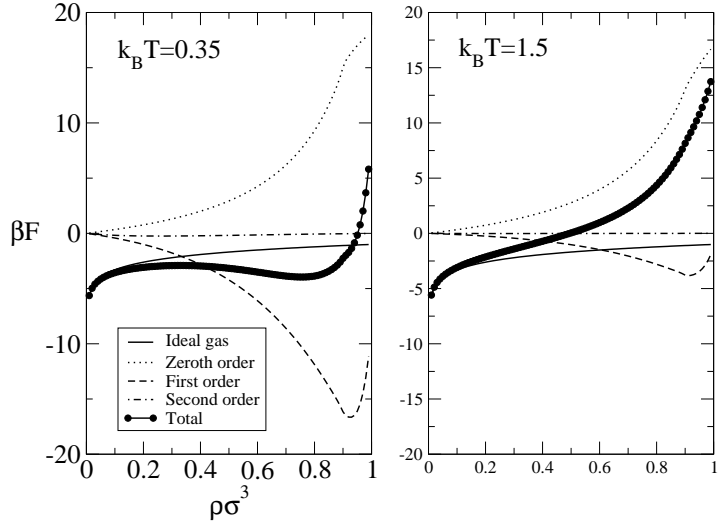}
\caption{The various terms contributing to the total free energy as a
function of density for two different temperatures. At the lower
temperature, the first order contribution dominates the hard-sphere
contribution whereas at higher temperatures, the zeroth order terms dominate.
}
\label{fig2}
\end{figure*}

In the case of the lower temperature, $k_{B}T/\varepsilon =0.35$, the first
order contributions quickly grow with density until at high densities, they
are larger than the zeroth order contributions thus suggesting that the
perturbation theory will not prove very accurate. At the higher temperature, 
$k_{B}T/\varepsilon =1.5$, the first order contributions are much better
controlled and we expect the perturbation theory to be relatively accurate.

We have also tested the various approaches to the selection of the
separation point of the potential - the WCA prescription, eq.(\ref{break1}),
the WCAR prescription and minimization of the free energy with respect to $%
r_{0}$. As expected, the only significant differences occur at high density,
where variations of the free energy of 10\% occur, but we find virtually no
effect on the phase diagram.

\section{Phase diagrams}

Figure 3 shows the phase diagram as calculated from the WCAR theory 
for the potential and parameters used by ten Wolde
and Frenkel\cite{tWF} and its comparison to the results of Monte Carlo
simulations by these authors for $\alpha=50$. The lines, from our calculations, and the symbols,
from the simulations, divide the density-temperature phase diagram into
three parts: the liquid region (low density and high temperature), the
fluid-solid coexistence region and the solid region (at high density). In
the calculations, the lines are determined by finding, for a given
temperature, the liquid and solid densities that give equal pressures and
chemical potentials for the two phases as determined using eq.(\ref{Pmu})
based on the liquid and solid free energy calculations (which differ only in
the equation of state and pair distribution function of the reference
states).  The fluid-fluid coexistence is determined similarly except that the
free energy for both phases is calculated using the same reference state
(the hard-sphere fluid) with the resulting free energy exhibiting a Van der
Waals loop.

\begin{figure*}[tbp]
\includegraphics[angle=0,scale=0.4]{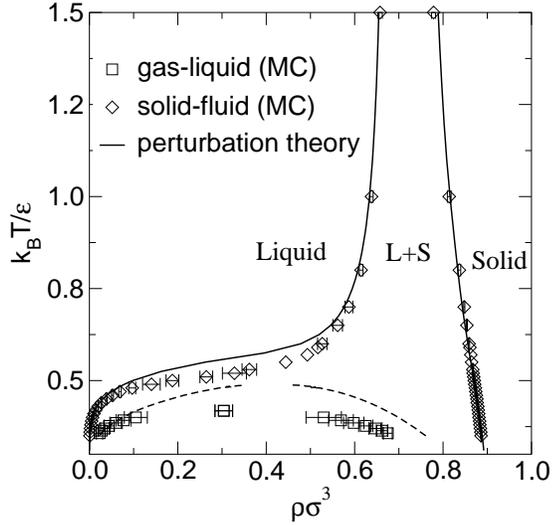}
\caption{Comparison of the predicted phase diagram, lines, to the Monte
Carlo results, symbols, of ref.\protect\cite{tWF} for $\alpha=50$. Some error bars are
superposed on the symbols.}
\label{fig3}
\end{figure*}

The calculations and simulations are in good qualitative agreement with a
fluid-fluid critical point that is suppressed by the fluid-solid phase
boundaries. The values of the coexisting densities are in good agreement at
low temperatures, where the liquid density is very low and at high
temperatures. That these limits agree is as expected from our discussion of
the relative sizes of the various contributions to the free energy. It is
perhaps surprising that the agreement is so good even for temperatures as
low as $k_{B}T/\varepsilon \sim 1$. The intermediate temperature values,
where the attractive tail and finite density effects are important, are the
most poorly described. The same is true of the fluid-fluid coexistence
curve. The critical point is estimated to occur at about $k_{B}T/\varepsilon
\sim 0.48$ and $\rho \sigma ^{3}\sim 0.4$ whereas the simulation results are 
$k_{B}T/\varepsilon \sim 0.4$ and $\rho \sigma ^{3}\sim 0.3$. We have tested
these results by using different choices for the pair distribution function
of the reference state (see eqs.(\ref{g1})-(\ref{g2})), and different
choices for the division of the potential (such as minimizing the free
energy with respect to the break point) but none of these alternatives
produces any significant change.

An interesting feature of short-ranged interactions is that under some
circumstances, they give rise to solid-solid transitions where the lattice
structure remains the same but solids of different densities can coexist
(i.e. a van der Waals loop occurs in the solid free energy)\cite%
{BausSolidSolid}. We have searched for,  but find no evidence of, such a
transition with the present potential.

\begin{figure*}[tbp]
\includegraphics[angle=0,scale=0.4]{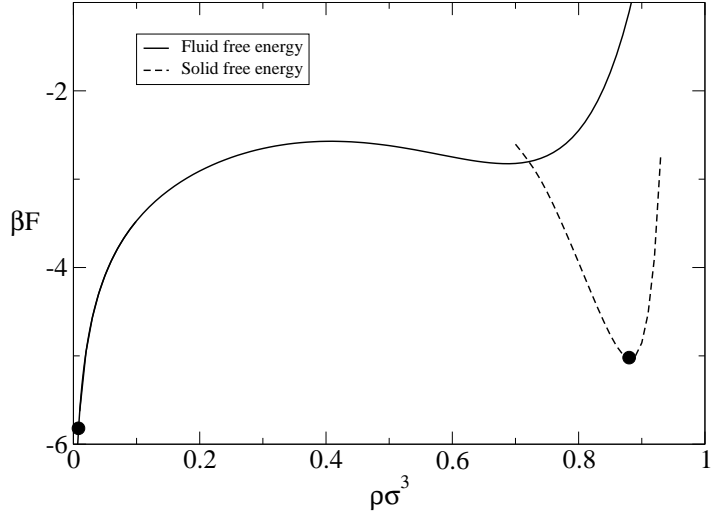}
\caption{Calculated free energies as a function of density for the liquid
and solid phases at $k_{B}T/\protect\epsilon = 0.4$ and for $\protect\alpha %
= 50$.The points mark the location of the coexisting phases.}
\label{fig5}
\end{figure*}

To give some idea of the typical energy barrier between the coexisting
phases, we show in Fig. \ref{fig5} the calculated isothermal free energies
as a function of density between the coexisting fluid and solid phases at $%
k_{B}T/\epsilon = 0.4$ for the short-ranged ($\alpha = 50$) potential. The
fluid has a density of 0.008 and Helmholtz free energy of -5.82 in reduced
units. The maximum free energy is -2.57 and the solid free energy is -5.02
at a density of 0.88.

\begin{figure*}[tbp]
\includegraphics[angle=0,scale=0.4]{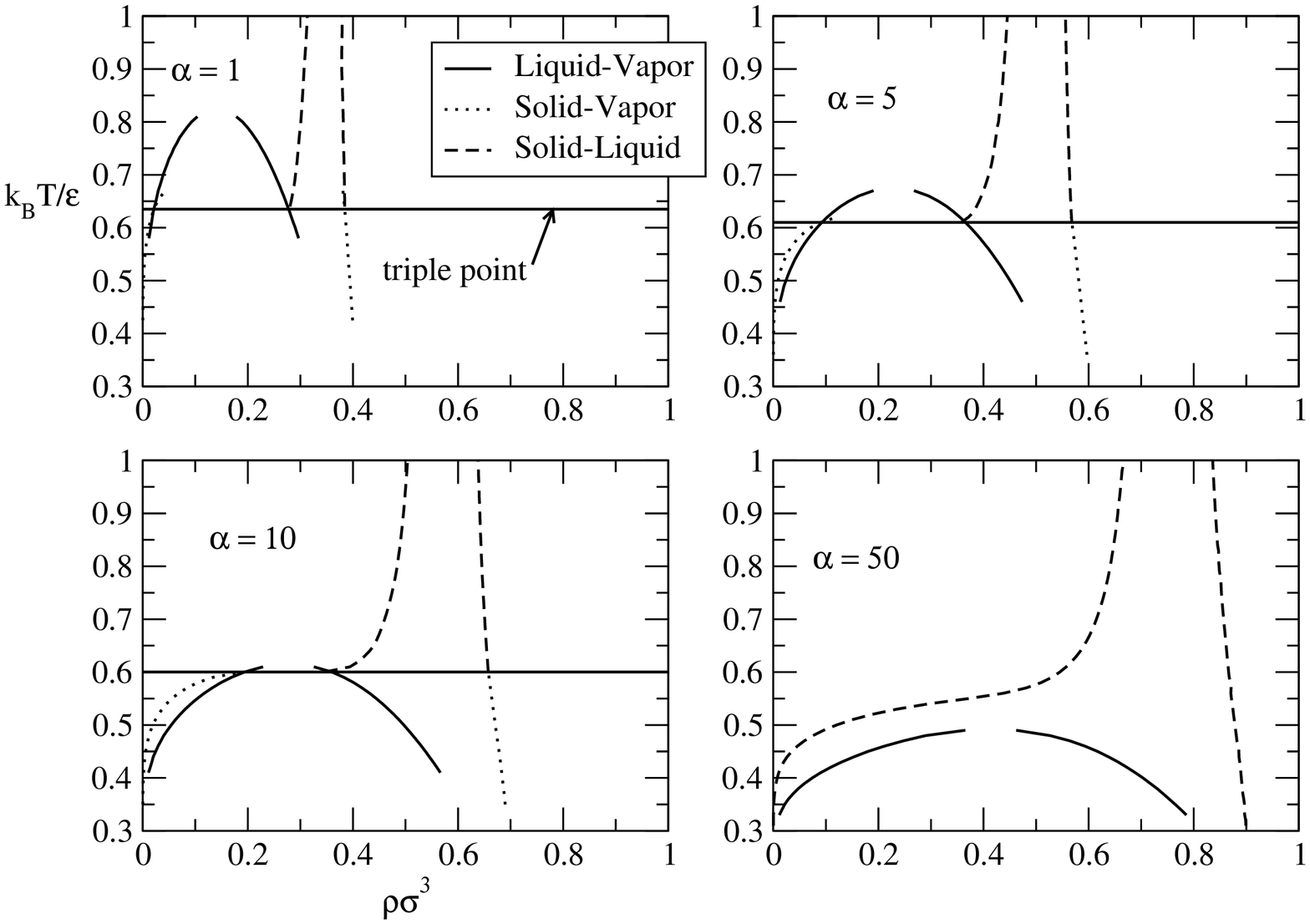}
\caption{Calculated phase diagrams as a function of $\protect\alpha$ showing
that the critical point is suppressed for $\protect\alpha \ge 10$.}
\label{fig4}
\end{figure*}

Figure \ref{fig4} shows the phase diagrams calculated from the WCAR theory
as a function of the range of the potential (i.e., different values of $%
\alpha $). For $\alpha=1$, for which the minimum of the potential well is $%
r_{min}=1.5$ and corresponding to a tail that closely resembles a standard
Lennard-Jones interaction, the phase diagram has the classical form
exhibiting three stable phases, a critical point and a triple point. As $%
\alpha$ increases, and the range of the potential decreases, the critical
point moves towards the triple point. Even for $\alpha=5$ and $r_{min}=1.31$%
, the critical point lies very near the triple point and the two become
nearly identical for $\alpha=10$ and $r_{min} = 1.26$. Our conclusion is
that for this model, the suppression of the triple point occurs when the
range of the potential, as characterized by its minimum, falls to about a
quarter of the hard-core diameter.

\section{Conclusions}

Our aim here has been to provide a fundamental model of protein crystallization without the need
for parameterizations other than the interaction potential. Since the
potential for globular proteins can be tuned, by varying e.g. the background
ionic strength of the solutions, this provides a rather direct connection
between theoretical indications of  favorable conditions for nucleation and
experimentally accessible control parameters.

We have shown that thermodynamic perturbation theory gives a good,
semi-quantitative estimate of the phase diagram of a model interaction for
globular proteins. The accuracy of the perturbation theory is expected to
improve as the range of the potential increases so, e.g., the prediction of
the value of $\alpha $ at which the critical point becomes suppressed is
expected to be reasonably accurate.  Unlike the results of a recent study of
colloids interacting via short-ranged potentials\cite{HansenPert}, we do not
find that the second order terms in the high-temperature expansion play an
important role in the structure of the phase diagram. 

This free energy calculation, which only uses the interaction model as
input, should be contrasted with other more phenomenological approaches. In
phase field models, the free energy is taken to be a function of one or more
order parameters. The actual form of the free energy is typically of the
Landau form which is to say, a square-gradient term plus an algebraic
function of more than second order in the order parameter. The coefficients
of these terms must be fitted to experimental data and the adequacy of the
assumed function is difficult to assess. Similarly, the recent density
functional models of Talanquer\cite{OxtobyProtein} and Shiryayev\cite%
{GuntonProtein} depend on an ad hoc free energy functional, based on the van
der Waals free energy model for the fluid, with several phenomenological
parameters.

We believe that our work can serve as the basis for further theoretical
study of the nucleation of globular proteins using density functional
theory. While the present description of the two phases requires as input
separate equations of state and pair distribution functions for the
reference hard sphere fluid and solid phases, standard methods exist for
interpolating between these so as to provide a single, unified free energy
functional suitable to the study of free energy barriers (see,e.g. ref.\cite%
{OLW}). Such a unified model can be used to study static properties, such as
the structure of the critical nucleas, using density functional theory as
well as the effect of fluctuations on the transition rates by the  addition of
noise obeying  the  fluctuation-dissipation theorem.

Finally, it would be desirable to confront the approach developed here to experiments
aiming to determine the interaction potential and the phase diagram of concrete globular
proteins of interest such as lysozyme and catalase. In recent years, considerable effort was devoted to
protein crystallization under microgravity conditions on the grounds that some undesirable effects such as density
gradients and advection present in earth-bound experiments can be virtually suppressed\cite{GarciaRuiz}.
In parallel, earth-bound experiments are being carried out to determine conditions and parameters to be used in 
a microgravity experiment.  In either case, the role of the metastable critical point has so far not
been addressed in a detailed manner. We believe that the availability of a theory as parameter-free as possible
like the one developed in the present work could provide the frame for undertaking such a study on a
rational basis. 

\begin{acknowledgements}
It is our pleasure to thank Pieter ten Wolde and
Daan Frenkel for making their simulation results available to us.
We have benefited from discussions with Ingrid Zegers and Vassilios Basios.
 This work was supportd in part by the European Space Agency
under contract number C90105.
\end{acknowledgements}

\bigskip

\appendix

\section{Evaluation of long-ranged contribution to the free energy}

We begin by writing the first order contribution of the long-ranged
potential as%
\begin{equation}
\int d\overrightarrow{r}\;g_{hs}\left( r\right) w\left( r\right) =\int d%
\overrightarrow{r}\;g_{hs}\left( r\right) v\left( r\right) -\int d%
\overrightarrow{r}\;g_{hs}\left( r\right) v_{0}\left( r\right) 
\end{equation}%
so that the second term involves the very short ranged function $v_{0}\left(
r\right) $ and is easily performed numerically. Our focus is therefore on
the evaluation of the first term on the right. If we write the potential as
the sum of a hard-core and a continuous tail 
\begin{equation}
v\left( r\right) =v_{hs}\left( r\right) +\Theta \left( r-\sigma \right)
v_{tail}\left( r\right) 
\end{equation}%
and the effective hard-sphere diameter $d$ $\geq $ $\sigma $, as it
clearly will always be, then 
\begin{eqnarray}
\int d\overrightarrow{r}\;g_{hs}\left( r\right) v\left( r\right)  &=&\int d%
\overrightarrow{r}\;\Theta \left( r-d\right) y_{hs}\left( r\right) v\left(
r\right)  \\
&=&\int d\overrightarrow{r}\;\Theta \left( r-d\right) y_{hs}\left( r\right)
v_{tail}\left( r\right)   \nonumber \\
&=&\int d\overrightarrow{r}\;g_{hs}\left( r\right) v_{tail}\left( r\right)  
\nonumber
\end{eqnarray}%
so that we can ignore the discontinuity of the hard-core potential and treat
and simply deal with the continuous tail potential. The first term can be
evaluated by introducing the inverse Laplace transform of $rv_{tail}\left(
r\right) $ , 
\begin{equation}
rv_{tail}\left( r\right) =\int_{0}^{\infty }ds\;\exp \left( -sr\right)
V_{tail}\left( s\right) 
\end{equation}%
and likewise for $rg_{hs}\left( r\right) $ so that%
\begin{eqnarray}
\int d\overrightarrow{r}\;g_{hs}\left( r\right) v_{tail}\left( r\right) 
&=&4\pi \int_{0}^{\infty }dr\;r^{2}g_{hs}\left( r\right) v_{tail}\left(
r\right)   \label{lt} \\
&=&4\pi \int_{0}^{\infty }ds\;V_{tail}\left( s\right) \int_{0}^{\infty
}dr\;rg_{hs}\left( r\right) \exp \left( -sr\right)   \nonumber \\
&=&4\pi \int_{0}^{\infty }ds\;V_{tail}\left( s\right) G\left( s\right)  
\nonumber
\end{eqnarray}%
where $G(s)$ is the Laplace transform of $rg_{hs}\left( r\right) $, which is
known analytic function in the PY approximation%
\begin{eqnarray}
G(s;d) &=&d^{2}G_{PY}(sd) \\
G_{PY}(x) &=&\frac{x\exp \left( -x\right) F\left( x\right) }{1+12\eta \exp
\left( -x\right) F\left( x\right) }  \nonumber \\
F\left( x\right)  &=&-\frac{1}{12\eta }\frac{1+Ax}{1+\left( A-1\right)
x+\left( \frac{1}{2}-A\right) x^{2}+\left( \frac{1}{2}A-\frac{1+2\eta }{%
12\eta }\right) x^{3}}  \nonumber \\
A &=&\frac{1+\eta /2}{1+2\eta }.  \nonumber
\end{eqnarray}%
The integral in eq.(\ref{lt}) is controlled by the exponential decay of $%
G(s;d)$ and is easily performed numerically. Note that for the ten
Wolde-Frenkel potential, we have that%
\begin{equation}
V_{tail}\left( s\right) =\frac{\varepsilon }{960\alpha ^{2}}\left( 
\begin{array}{c}
\left( \left( s\sigma \right) ^{5}+45\left( s\sigma \right) ^{3}+\left(
105+480\alpha \right) s\sigma \right) \sinh s\sigma  \\ 
-\left( 10\left( s\sigma \right) ^{4}+\left( 105+480\alpha \right) \left(
s\sigma \right) ^{2}\right) \cosh s\sigma 
\end{array}%
\right) .
\end{equation}

The Percus-Yevick pair distribution function becomes exact at low densities
but is only semi-quantitatively accurate at moderate to high densities.
Compared to the pdf determined from computer simulations, its oscillations
are slightly out of phase and the pressure calculated from it is in error.
The Verlet Weiss pair distribution function is a semi-empirical modification
of the basic Percus-Yevick result designed to correct these flaws. It is
written as%
\begin{equation}
g_{VW}\left( r;\rho ,d\right) =\Theta \left( r-d\right) \left( g_{PY}\left(
r;\rho ,d_{0}\right) +\frac{C}{r}\exp \left( -m(r-d)\right) \cos \left(
m\left( r-d\right) \right) \right)   \label{vw1}
\end{equation}%
where the step function $\Theta \left( r-d\right) $ ensures the fundamental
property that the pdf vanishes inside the core, $d_{0}$ is an effective
hard-sphere diameter which has the effect of shifting the phase of the
oscillations, and $C$ and $m$ are chosen to give the accurate
Carnahan-Starling equation of state via both the pressure equation and the
compressibility equation. To apply the Laplace technique in this case
requires some care since what we know is the Laplace transform of $%
g_{PY}\left( r;\rho ,d_{0}\right) $ and not that of $\Theta \left(
r-d\right) g_{PY}\left( r;\rho ,d_{0}\right) $. So we rewrite eq.(\ref{vw1})
as 
\begin{eqnarray}
g_{VW}\left( r;\rho ,d\right)  &=&g_{PY}\left( r;\rho ,d_{0}\right) +\left(
\Theta \left( r-d\right) -\Theta \left( r-d_{0}\right) \right) g_{PY}\left(
r;\rho ,d_{0}\right)  \\
&&+\Theta \left( r-d\right) \frac{C}{r}\exp \left( -m(r-d)\right) \cos
\left( m\left( r-d\right) \right)   \nonumber
\end{eqnarray}%
thus separating out the known PY\ contribution. This gives%
\begin{eqnarray}
\int d\overrightarrow{r}\;g_{hs}\left( r\right) v_{tail}\left( r\right) 
&=&\int d\overrightarrow{r}\;g_{PY}\left( r;\rho ,d_{0}\right)
v_{tail}\left( r\right)  \\
&&+4\pi \int_{d_{0}}^{d}r^{2}dr\;g_{PY}\left( r;\rho ,d_{0}\right)
v_{tail}\left( r\right)   \nonumber \\
&&+4\pi \int_{d}^{\infty }r^{2}dr\;\frac{C}{r}\exp \left( -m(r-d)\right)
\cos \left( m\left( r-d\right) \right) v_{tail}\left( r\right)   \nonumber
\end{eqnarray}%
where the first integral can be evaluated via the Laplace transform
technique, provided that $d_{0}\geq \sigma $, the second integral is over a
finite interval (for which one could analytically approximate the pair
distribution function as in ref.\cite{HendGrundke}) while the third
integral is easily evaluated numerically. All parts of the calculation are
therefore well controlled.

Finally, we note that the same techniques can be adapted to the evaluation
of the second order contribution to the free energy.

\bigskip

\bibliographystyle{prsty}
\bibliography{dft}

\bigskip

\end{document}